\documentclass[aps,twocolumn,showpacs,floatfix,groupedaddress,amsmath,amssymb,prl]{revtex4-1}
\usepackage{graphicx}
\usepackage{multirow}
\usepackage{color}
\usepackage[]{qcircuit}

\usepackage{xfrac}

\usepackage{array}
\usepackage{hyperref}
\newcommand{\Tr}[1]{\operatorname{Tr}( #1 )}
\newcommand{\ket}[1]{\lvert #1 \rangle}
\newcommand{\bra}[1]{\langle #1 \rvert}

\begin{document}
\title{A geometric pathway to scalable quantum sensing}
\author{Mattias T. Johnsson, Nabomita Roy Mukty, Daniel Burgarth, Thomas Volz}

\author{Gavin K. Brennen}
\email{gavin.brennen@mq.edu.au}

\affiliation{Center for Engineered Quantum Systems, Dept. of Physics \& Astronomy, Macquarie University, 2109 NSW, Australia}

\begin{abstract}
Entangled resources enable quantum sensing that achieves Heisenberg scaling, a quadratic improvement on the standard quantum limit, but preparing large $N$ spin entangled states is challenging in the presence of decoherence. We present a quantum control strategy using highly nonlinear geometric phase gates which can be used for generic state or unitary synthesis on the Dicke subspace with $O(N)$ or $O(N^2)$ gates respectively. The method uses a dispersive coupling of the spins to a common bosonic mode and does not require addressability, special detunings, or interactions between the spins. By using amplitude amplification our control sequence for preparing states ideal for metrology can be significantly simplified to $O(N^{5/4})$ geometric phase gates with parameters that are more robust to mode decay. The geometrically closed path of the control operations ensures the gates are insensitive to the initial state of the mode and the sequence has built-in dynamical decoupling providing resilience to dephasing errors. 
%We present a quantum control strategy for preparing Dicke states on spin ensembles for use in precision metrology. The method uses a dispersive coupling of $n$ spins to a common bosonic mode and does not require selective addressing, adiabatic state transfer, or direct interactions between the spins. Using a control sequence based on amplitude amplification, a target state can be prepared using $O(n^{5/4})$ geometric phase gates. Due to the geometrically closed path of the control operators on the joint mode-spin space, the sequence has dynamical decoupling built in providing resilience to dephasing errors.
\end{abstract}

\maketitle

{\it {Introduction}}.---
Quantum enhanced sensing offers the possibility of using entanglement in an essential way to measure fields with a precision superior to that which can be obtained with unentangled resources \cite{wasilewski2010,taylor2008, wang2019,giovannetti2011}. Entangled resources allow the measurement sensitivity to scale as $1/N$ with respect to the resources applied (so-called Heisenberg scaling), compared to the $1/\sqrt{N}$ obtained otherwise (the standard quantum limit, or shot-noise limit) \cite{giovannetti2004, pirandola2018,giovannetti2011}.

Creating large-scale entanglement in multipartite systems for the purposes of metrology is a difficult problem for a number of reasons. There is the difficulty in precisely constructing the required quantum state using realistic quantum operations, the need to protect that quantum state from decoherence and loss \cite{Demkowicz-Dobrzanski:2012xq}, and the problem of carrying out a number of quantum measurement operations on the state with precise control.

From a metrology perspective, there is also the issue that many schemes claim to achieve Heisenberg limit by virtue of quadratic scaling of the Fisher information of the system \cite{zwierz2010}. While this ensures that there is an observable which has an uncertainty that scales as $1/N$ with respect to some resource, it does not specify what that observable is, or require it be a convenient experimentally measurable quantity.
%\begin{figure}[ht!]
%\begin{center}
%\includegraphics[scale=0.3]{Cartoon.pdf}
%\end{center}
%\caption{Illustration of the state preparation
%protocol. By attaching a bosonic mode dispersively coupled to a set of $N$ spins, geometric phase gates and global rotations drive a system to an entangled state ready for use in quantum sensing. The gate control sequence forming the parallelogram in the mode phase space is shown, as are the times where dynamical decoupling is applied.
%}
%\label{fig:cartoon}
%\end{figure}
There have been attempts to address these problems in various ways. For example, to mitigate the decoherence issue, recent work has suggested using quantum error correction assisted metrology (see \cite{Zhou:2018am} and references therein) or phase protected metrology \cite{Bartlett_2017}. Such workarounds require the ability to perform complex quantum control in the former case or engineered interactions in the latter.

Here we present a state preparation scheme and measurement protocol using geometric phase gates generalizing Ref.~\cite{Wang:02}. Our protocol addresses the issues of state preparation, decoherence protection, and choice of measurement operator.
%(shown schematically in Fig.~\ref{fig:cartoon}).
It is relatively simple to engineer as it involves only the coupling of an ensemble of qubits to a common bosonic mode, e.g. a cavity or mechanical oscillation, as well as simple global control pulses on the spins and mode. As such it is adaptable to a variety of architectures at the forefront of quantum control including NV centres in diamond, trapped ion arrays, Rydberg atoms, and superconducting qubits.

Unlike previous work our scheme does not require special engineering of the physical layout of the spins, special detunings for adiabatic state preparation, addressability, or direct interaction between the spins. Furthermore, it exceeds the performance of spin squeezing protocols because of the highly nonlinear nature of the geometric phase gates used. Another advantage is that due to the geometric nature of the gate, it is completely insensitive to variations or uncertainties in the rate at which the perimeter is traversed.

 Furthermore, our protocol has dynamical decoupling built in, providing resilience against dephasing during the state preparation, which is the dominant source of noise in many physical implementations. While dynamical decoupling has been considered \cite{Xu2004} in the context of the M{\o}llmer-S{\o}renson geometric gate \cite{MS1999}, our scheme extends this to a highly nonlinear geometric phase gate and a full quantum state preparation algorithm.
 
 %For plausible assumptions on the form of the system-bath spectral density, we obtain a suppression on the dephasing rate of two orders of magnitude.

{\it Method}.---
We consider a collection of two-level spin half systems, and define the collective raising and lowering angular momentum operators as
$J^+=\sum_{j=1}^N \sigma^+_j, J^-=(J^+)^{\dagger}$, and the components of the total angular momentum vector are $J^x=(J^++J^-)/2, J^y=(J^+-J^-)/2i, J^z=\sum_j (\ket{0}_j\bra{0}-\ket{1}_j\bra{1})/2$. Dicke states are simultaneous eigenstates of angular momentum $J$ and $J^z$: $\ket{J=N/2,J^z=M}$, $M=-J,\ldots, J$. 
%Transition rates between adjacent states in the Dicke ladder are:
%\begin{equation*}
%\Gamma_{M\rightarrow M\pm 1}=\Gamma \bra{J,M}J^{\mp}J^{\pm}\ket{J,M}=\Gamma (J\mp M)(J\pm M+1),
%\end{equation*}
%where $\Gamma$ is the single spin decay rate. At the middle of the Dicke ladder (near $M=0$), these rates are $O(N)$ times faster than for $N$ independent spins and the Dicke state $\ket{J,0}$ is referred to as superradiant.
%when emitting or, in the reciprocal process, as superabsorptive. By suitable reservoir engineering, superabsorption can be exploited for photon detection and energy harvesting \cite{Higgins:2014qq}.

Consider the measurement of a field which generates a collective spin rotation about an axis perpendicular to $\hat{z}$ given by $U(\eta)=\exp \left[ i\eta(J^x \sin \delta + J^y\cos\delta) \right]$. For a measurement operator $O$ on the system, the single shot estimation of the parameter $\eta$ has variance
\begin{equation}
(\Delta \eta)^2=\frac{(\Delta O(\eta))^2}{|\partial_{\eta}\langle O(\eta)\rangle|^2}.
\label{eqDeltaBetaDefinition}
\end{equation}
When the measured observable is $O={J^z}^2$, the parameter variance is \cite{Apellaniz_2015}
\begin{align*}
(\Delta \eta)^2 & = \big((\Delta {J^x}^2)^2 f(\eta)+4\langle {J^x}^2\rangle-3\langle {J^y}^2\rangle-2\langle {J^z}^2\rangle  \\
& \!\! \times(1+\langle {J^x}^2\rangle)+6\langle J^z{J^x}^2J^z\rangle\big)(4(\langle {J^x}^2\rangle-\langle {J^z}^2\rangle)^2)^{-1}
\end{align*}
with $f(\eta)=\frac{(\Delta {J^z}^2)^2}{(\Delta {J^x}^2)^2\tan^2(\eta)}+\tan^2(\eta)$.
When the initial state is the Dicke state $\ket{J,0}$, the estimate angle satisfies $\sin^2(\eta)=\frac{8\langle {J^z}^2(\eta)\rangle}{N(N+2)}$ and 
the uncertainty in the measured angle is minimized at $\eta_{\rm min}=0$ such that the quantum Cram\'er-Rao bound is saturated: 
\begin{equation}
  (\Delta \eta)^2=\frac{2}{N(N+2)}.  
  \label{QCRB}
\end{equation}
Note it is not essential that we know the angle $\delta$ of the field direction in the $\hat{x}-\hat{y}$ plane \cite{Note2}.
%. To see this, suppose a field is aligned with an angle $\delta$ in the $\hat{x}-\hat{y}$ plane, meaning the unitary is given by
%\begin{eqnarray}
%    U(\eta) &=& \exp \left[ i\eta(J^x \sin \delta + J^y\cos\delta) \right] \nonumber \\
%            &=& \exp (i\delta J^z) \exp(i\eta J^y) \exp(-i\delta J^z).
%\end{eqnarray}
%We now measure the variance of our observable ${J^z}^2$ on our initial state $|J, M=0\rangle$ as before, with
%\begin{equation}
%(\Delta J^{z2})^2 = \langle {J^z}^4 \rangle - \langle {J^z}^2 \rangle^2
%\end{equation}
%where for any power $s$
%\begin{equation}
%\langle {J^z}^s \rangle = \langle J, J^z=0 | U^{\dagger} (\eta) \, {J^z}^s \, U(\eta) | J, J^z=0 \rangle
%\end{equation}
%Our measured observable is $O={J^z}^2$, and the associated precision is given by Eq.~(\ref{eqDeltaBetaDefinition}), independent of $\delta$.

The best known quantum algorithm for deterministically preparing a Dicke state $\ket{J,M}$ requires $O((N/2+M)N)$ gates and has a circuit 
depth $O(N)$ \cite{2019arXiv190407358B}. This works even for a linear nearest neighbour architecture, but requires a universal gate set and full addressability.
Other proposals exist \cite{PhysRevA.95.013845,Higgins:2014qq,keatingET2016,HumeET2009}, but they all suffer from drawbacks such as not scaling beyond a few spins, strong adiabaticity or geometry, contraints, requiring large initial Fock states of motional modes, and couplings causing transitions outside the Dicke space.

%Another proposal avoids addressability in the ultra-strong coupling regime of circuit QED systems \cite{PhysRevA.95.013845} via selective resonant interactions at different couplings in order to transfer excitations one by one to the spin ensemble, but is difficult to scale up while satisfying the large detuning constraint required. Another strategy is to use interactions between the spins for state preparation.  In the proposal of Ref. \cite{Higgins:2014qq}, dipole-dipole interactions provide a nonlinear energy shift in the Dicke ladder, allowing Dicke state preparation using chirped excitation pulses and/or measurement and feedback control. However, the dipole-dipole interaction creates transitions outside the Dicke space, and resolving transitions for a large number of spins is challenging. The use of Rydberg systems has also been proposed \cite{keatingET2016}, but this scheme involves control pulses found numerically, and even in the absence of noise and dephasing is limited to $\sim 10$ spins. It addition suffers from control noise and also requires satisfying an adiabaticity criterion with unclear scaling.

%In contrast, our geometric phase gate (GPG) based approach for preparing Dicke states has depth $O(N^{5/4})$ and requires no direct coupling between spins, no addressability, and uses only global rotations and semi-classical control on an external bosonic mode with no special field detunings required.

%
%\subsection{Grover search method}

In our setup 
%(see Fig.~\ref{fig:cavity_and_gate}) 
we assume $N$ spins with homogeneous energy splittings described by a free Hamiltonian $H_0=\omega_0 J^z$ 
%(setting $\hbar\equiv 1$),
which can be controlled by semi-classical fields performing global rotations generated by $J^x,J^y$. Additionally, we assume the ensemble is coupled to a single quantized bosonic mode $\hat{a}$.
Our scheme requires a dispersive coupling between the $n$ spins and the bosonic mode of the form $V=g a^{\dagger}a J^z$.
%We assume $g>0$ but the case $g<0$ follows easily as described below. 
%Such dispersive interactions have already been demonstrated with trapped ions coupled to an optical cavity \cite{PhysRevLett.122.153603}, and Rydberg atoms or superconducting qubits coupled to microwave resonators \cite{Sayrin:2011eu, Wang1087}.  
%By complementing this interaction with field displacement operators on a quantized bosonic mode it is possible to generate a GPG which can produce many body entanglement between the spins while in the end being disentanged from the mode.

The geometric phase gate (GPG) makes use of two basic operators \cite{Wang:02,Jiang:2008jo}, the displacement operator $D(\alpha)=e^{\alpha a^{\dagger}-\alpha^{\ast}a}$ and the rotation operator 
$R(\theta J^z)=e^{i\theta  a^{\dagger}a J^z}$
which perform a closed loop in the mode phase space 
%(see Fig.~\ref{fig:cavity_and_gate}(b)):
\begin{eqnarray} 
U_{GPG}(\theta,\phi,\chi)&=&D(-\beta)R(\theta J^z)D(-\alpha)R(-\theta J^z) \nonumber \\
&&\times \,\, D(\beta)R(\theta J^z)D(\alpha)R(-\theta J^z) \nonumber \\
&=&e^{-i 2 \chi\sin(\theta J^z+\phi)}.
\label{seq}
\end{eqnarray}
where $\phi=\arg(\alpha)-\arg(\beta)$ and $\chi=|\alpha\beta|$.

The system and the mode are decoupled at the end of GPG cycle. Also, if the mode begins in the vacuum state, it ends in the vacuum state and the first operation $R(-\theta J^z)$ in Eq.~(\ref{seq}) is not needed. 
%However, as explained below it can be useful to include the first step as free evolution, in order to negate the total free evolution and to suppress dephasing errors. 
In the GPG it is necessary to evolve by both $R(\theta J^z)$ and $R(-\theta J^z)$. This can be done by conjugating with a global flip of the spins $R(\theta J^z)=e^{-i\pi J^x}R(-\theta J^z)e^{i\pi J^x}$, implying that the GPG can be generated regardless of the sign of the dispersive coupling strength $g$. Furthermore, because $R(\pm\theta J^z)$ commutes with $H_0$ at all steps, this conjugation will cancel the free evolution accumulated during the GPG.
%If the displacement operators are fast compared to $1/\omega_0,1/g$ then the total time for the GPG is $t_{GPG}=4\theta/g$.
  
%We assume the number of spins $n$ is even, although the protocol can easily be adapted to prepare Dicke states for odd $N$ as described below. 
Assuming the number of spins $n$ is even, we consider $N/2$ sequential applications of the GPG (see also \cite{Wang:02}):
\begin{eqnarray*}
W(\ell)&=&\prod_{k=1}^{N/2}U_{GPG}(\theta_k,\phi_k(\ell),\chi) \nonumber \\
&=&\sum_{M=-J}^J e^{-i2\chi\sum_{k=1}^{N/2} \sin(\theta_k M+\phi_k(\ell))}\ket{J,M}\bra{J,M},
\end{eqnarray*}
with $\ell=0,\ldots, N$.
\begin{equation*}
\theta_k=\frac{2\pi k}{N+1},\quad \phi_k(\ell)=\frac{2\pi k (N/2-\ell)}{N+1}+\frac{\pi}{2},\quad \chi=\frac{\pi}{N+1},
\end{equation*}
%\begin{equation}
%\frac{2}{N+1}\sum_{k=1}^{N/2} \cos(\frac{2\pi k (M+N/2-\ell) }{N+1})=\delta_{\ell,M+N/2}-\frac{1}{N+1},
%\label{sumcos}
%\end{equation}
gives $W(\ell)=e^{-i\pi \ket{J,\ell-N/2}\bra{J,\ell-N/2}} $
meaning it applies a $\pi$ phase shift on the symmetric state with $\ell$ excitations. For $N$ odd
%we have the sum
%\[
%\frac{1}{N+1}\sum_{k=1}^{N} \cos(\frac{2\pi k (M+N/2-\ell) }{N+1})=\delta_{\ell,M+N/2}-\frac{1}{N+1},
%\]
%so
we can use $N$ GPGs with the same angles $\theta_k,\phi_k(\ell)$ as above but with $\chi=\frac{\pi}{2(N+1)}$.

Given the control toolbox above of global rotations and the GPGs, one can synthesize an arbitrary unitary operator on the Dicke subspace. Writing  $U=\sum_{k=1}^{N}e^{i\lambda_k\ket{\lambda_k}\bra{\lambda_k}}$, where $\{\ket{\lambda_k}\}$ form an orthonormal basis on the Dicke subspace, and $\lambda_k\in \mathbb{R}$ (note since the global phase is irrelevant we have set $\lambda_{2J+1}=0$). This unitary can be decomposed as $U=\prod_{k=1}^{N-1}[K(\lambda_k)e^{i\lambda_k \ket{J,-J}\bra{J,-J}}K^{\dagger}(\lambda_k)]$, where 
$K(\lambda_k)$ is any unitary extension of the state synthesis mapping $\ket{J,-J}\rightarrow \ket{\lambda_k}$. The phasing unitary is the same as $W(0)$ but with $\chi\rightarrow \lambda_k/(N+1)$. To construct $K(\lambda_k)$, we find the decomposition:
$\tilde{K}=[\prod_{s=1}^{N-1}e^{i \beta_s J^y}U_{GPG}(\theta_s,\phi_s,\chi_s)]e^{-i J^y \frac{\pi}{2}}U_{GPG}(\frac{\pi}{2},0,\frac{\pi}{4})e^{i J^y \frac{\pi}{2}}$
gives very accurate results when optimized over the $4N-4$ free parameters $\{\beta_s,\theta_s,\phi_s,\chi_s\}$ \cite{Note3}. The overall complexity in GPG count using this approach is $N$ for state synthesis and $5 N^2/2$ for unitary synthesis. 

While the general state synthesis approach above can be used for building Dicke states,  the $N-1$ action angles $\{\chi_s\}$ that optimize state fidelity are $O(1)$ and this has implications for noise due to mode decay as described below. We now describe a way, based on amplitude amplification, to improve matters by only using the GPGs that appear in $W(\ell)$ that have action angles $O(1/N)$, and with the added advantage of providing an analytical solution to the Dicke state synthesis problem.
Starting with all spins down, i.e. in $\ket{J,-J}$, let the target state be $\ket{w}=\ket{J,0}.$ We will make use of the operators $U_w=e^{-i\pi \ket{w}\bra{w}}=W(N/2)$ and $U_s=e^{-i\pi \ket{s}\bra{s}}=e^{iJ^y\pi/2}W(0) e^{-iJ^y\pi/2}.$
In total the operators $U_w$ and $U_s$ each use $N/2$ GPGs.
The orbit of the initial state $\ket{s}$ under the operators $U_w$ and $U_s$, is restricted to a subspace spanned by the orthonormal states $\ket{w}$ and $\ket{s'}=\frac{\ket{s}-\ket{w}\bra{w}s\rangle}{\sqrt{1-|\bra{w}s\rangle|^2}}$, exactly as in Grover's algorithm.  
%Specifically, $U_w$ is a reflection across $\ket{s'}$ and $U_s$ is a reflection through $\ket{s}$ in this subspace exactly as in Grover's algorithm. 
The composite pulse is one Grover step 
$U_G=U_sU_w$. 
%Note, in between the steps $U_w$ and $U_s$, the accumulated drift from the free Hamiltonian $H_0$, which commutes with the GPGs, can be negated by a composite pulse sequence of 3 global rotation pulses via Euler decomposition.
Geometrically, relative to the state $\ket{s'}$, the initial state $\ket{s}$ is rotated by an angle $\delta/2$ toward $\ket{w}$, where $\delta=2 \sin^{-1}(|\bra{w}s\rangle|),$ and after each Grover step is rotated a further angle $\delta$ toward the target. The optimal number of Grover iterations to reach the target is 
$\#G=\Big\lfloor{\frac{\pi}{4|\bra{w}s\rangle|}}\Big\rfloor$ where the relevant overlap is $\bra{w}s\rangle =d^J_{-J,0}(\pi/2)  =2^{-J}\sqrt{(2J)!}/{J!}$.
%\begin{equation}
%\bra{w}s\rangle =d^J_{-J,0}(\pi/2)  =2^{-J}\sqrt{(2J)!}/{J!},
%\end{equation}
%where $d^{J}_{M',M}(\theta)=\bra{J,M'}e^{-iJ^y\theta}\ket{J,M}$ are the Wigner (small) d-matrix elements. 
For 
$J\gg 1$
%, using $x!\approx x^x e^{-x}\sqrt{2\pi x}$,
we have $\bra{w}s\rangle\approx (\pi J)^{-1/4}$. Then the optimal number of Grover steps is
\begin{equation}
 \#G=\lfloor \pi^{5/4}N^{1/4}/2^{9/4} \rfloor.
 \label{Tcount}
 \end{equation}
The fidelity overlap of the output state $\rho$ of the protocol with the target state is $F=\Tr{\ket{w}\bra{w} \rho}$. For the Grover method it is easily calculated as
\begin{equation}
F
%&=&|\bra{w}U_G^{\#G}\ket{s}|^2\\
=\sin^2((\#G+1/2)\delta) > 1-\sqrt{2/\pi N}.
\end{equation}
While the fidelity error falls off at least as fast as $\sqrt{2/\pi N}$ for all $N\gg 1$, if $N$ is near a value where the argument in Eq.~(\ref{Tcount}) is a half integer, i.e. $\lceil 32 (2k+1)^4/\pi^5\rceil$, with $k\in \mathbb{Z}$, the error will be much lower. For example, at $N=(10,70,260,700,1552)$ the fidelity error is $(1.84\times 10^{-4},1.57\times 10^{-5},1.68\times 10^{-6},3.65\times 10^{-8},1.92\times 10^{-8})$. The number of GPGs needed to prepare the Dicke state by the Grover method is $c\times N^{5/4}$ with a constant $c<1$.

\begin{figure}[tb]
\begin{center}
\includegraphics[width=8.6cm]{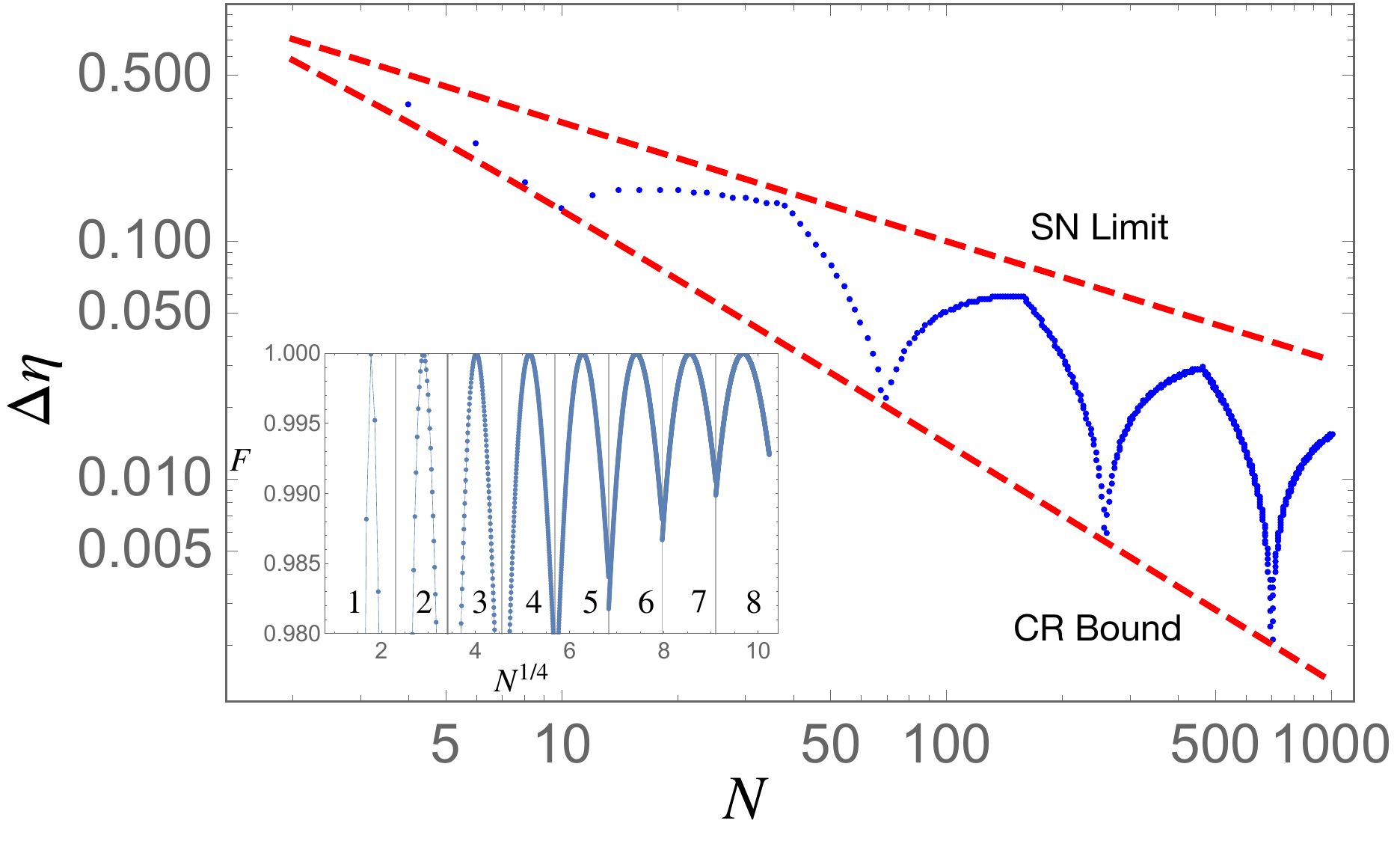}
\end{center}
\caption{Measurement precision $\Delta \eta$ as a function of number of spins (log-log scale). Shown are the shot noise limit, the quantum Cram{\'{e}}r-Rao bound Eq.~(\ref{QCRB}), and this protocol (blue). Inset: Fidelity for sets of ensemble sizes using the same number of Grover steps, which grow as $N^{1/4}$.}
\label{fig:DeltaEtaAsFunctionOfN}
\end{figure}

The effectiveness of our scheme when used for metrology is shown in Figure~\ref{fig:DeltaEtaAsFunctionOfN}, which shows the precision $\Delta \eta$ obtainable as a function of $N$, compared to that obtained from both the standard quantum limit and the ultimate Cram{\'{e}}r-Rao bound. It also shows the fidelity obtainable as a function of the number of spins $N$. The achievable fidelity is clearly optimized for specific spin values. 

%and each GPG has a dispersive interaction action angle of $\theta=g t=O(1)$, implying the total time for the state preparation is $O(N^{5/4}/g)$. 

We have focused on preparing the state $\ket{J,0}$, but with simple modifications our protocol can prepare any Dicke state $\ket{J,M}$. First use the initial state $\ket{s}=e^{i\epsilon_M J^y}\ket{J,-J}$, and second substitute the operators $U_w=W(M+N/2)$ and $U_s=e^{i\epsilon_M J^y}W(0)e^{-i\epsilon_M J^y}$ where $\epsilon_M=\cos^{-1}(M/J)$. Now the relevant overlap is $|\bra{w}s\rangle|=|d^{J}_{M,-J}(-\epsilon_M)|$, and for $J-|M|\gg 1$, $|d^{J}_{M,-J}(-\epsilon)|\approx (\sqrt{\pi J}\sin\epsilon_M)^{-1/2}$ \cite{doi:10.1063/1.1358305}, implying $\#G=O(N^{1/4})$.

Finally, measurement of ${J^z}^2$ could be done by again coupling the spins to the bosonic mode but now with a linear coupling $V_m=g_m J^z(a^{\dagger}+a)$ which generates a mode displacement depending on the collective spin projection. When the mode is in a number diagonal state, e.g. a thermal state, with mean excitation number $\bar{n}$, the measurement of $\hat{n}$ after a coupling time $\tau$ is $\langle \hat{n}\rangle=\bar{n}+(g_m \tau)^2 \langle {J^z}^2\rangle$.  
If mode number and quadratic spin operator measurements are difficult there are alternatives. One is estimation by a classical average over $p$ experiments: $E[\langle{J^z}^2\rangle]=\sum_{k=1}^p \frac{M(k)^2}{p}$, where $M(k)$ is the outcome of the $k$th measurement of $J^z$ \cite{Lucke:2011xd}. Another is, after accumulating the signal, one could invert the state preparation and measure $\ket{J,-J}\bra{J,-J}$ which gives the same precision scaling as Eq.~(\ref{eqDeltaBetaDefinition}), but doesn't require number resolved excitation counts.

There will be errors due to decay of the bosonic mode during the operations, as well as decoherence due to environmental coupling to the spins, which will degrade the fidelity. We now address these.  

{\it Mode damping}.---
We treat the mode as an open quantum system with decay rate $\kappa$. In order to disentangle the spins from the mode, the third and fourth displacement stages of the $k$-th GPG should be modified to $D(-\alpha_k)\rightarrow D(-\alpha_k e^{-\kappa \theta_k/g})$ and $D(-\beta_k)\rightarrow D(-\beta_k e^{-\kappa \theta_k/g})$. For simplicity we choose $|\alpha_k|=|\beta_k|$.
%\begin{figure}[tb]
%\begin{center}
%\includegraphics[width=8.6cm]{Figure4.pdf}
%\end{center}
%\caption{Performance of our protocol for preparing the Dicke state $\ket{J,0}$. (a) Probability distribution $P(M)$ in state $\ket{J,M}$ for the initial state $\ket{s}$ and the final state $U_G^2\ket{s}$ for $N=70$ spins after two Grover steps. The final fidelity error is $1-F=1-P_{\rm final}(0)=1.57\times 10^{-5}$. (b) Scalable performance at high fidelity. Sets of ensemble sizes using the same number of Grover steps, which grow as $N^{1/4}$, are indicated.}
%\label{fig:performance}
%\end{figure}
For an input spin state in the symmetric Dicke space $\rho=\sum_{M,M'}\rho_{M,M'}\ket{J,M}\bra{J,M'}$, the process for the $k-$th GPG with decay on the spins, including the modified displacement operations above, is \cite{Brennen:09}
\begin{equation*}
\begin{array}{lll}
\mathcal{E}^{(k)}(\rho)&=&U_{GPG}(\theta_k,\phi_k,\chi_k) \big[ \sum_{M,M'}R^{(k)}_{M,M'}\rho_{M,M'}.  \\
&&\ket{J,M}\bra{J,M'} \big] \times U_{GPG}^{\dagger}(\theta_k,\phi_k,\chi_k)
\end{array}
\end{equation*}
%We achieve best performance by choosing the displacement amplitudes according to
%\begin{equation}
%|\alpha_k\beta_k|=\frac{2\pi}{(n+1)(e^{-3\Gamma_{\rm gd}\theta_k/2g}+e^{-\Gamma_{\rm gd}\theta_k/2g})}.
%\end{equation}
where $\chi_k=|\alpha_k|^2 f(\theta_k)$, $f(\theta_k)=(e^{-3\theta_k\kappa/2g}+e^{-\theta_k\kappa/2g})/2$ and $\Gamma_{M,M'}$ and $\Delta_{M,M'}$ are given in the Supplementary Material \cite{refSupplemental}.

%\[
%R^{(k)}_{M,M'}=e^{-\Gamma_{M,M'}(\theta_k,\phi_k,\alpha_k)}e^{i\Delta_{M,M'}(\theta_k,\phi_k,\alpha_k)}
%\]
%with the factors $\Gamma_{M,M'}$ and $\Delta_{M,M'}$ given in the Supplementary Material.
The full operation is a concatenation of these imperfect processes $\mathcal{E}^{(k)}(\rho)$, and we characterise its accuracy with the process fidelity $F_{\mathrm{pro}}(\mathcal{E},U)$, which measures how close a quantum operation $\mathcal{E}$ is to the ideal operation $U$ \cite{nielsen_chuang}.

For each GPG we can readily find the lower bound on the process fidelity (see Supplementary Material \cite{refSupplemental})
\begin{equation*}
F_{\mathrm{pro}}(\mathcal{E},U_{\rm GPG})>e^{-6\pi \chi_k \kappa/f(\theta_k) g}\cos(\chi_k 4\pi \kappa/f(\theta_k) g).
\label{fullfid}
\end{equation*}
Note the scaling of the exponent with $1/N$ since the action angles $\chi$ are $O(1/N)$.
For the composite phasing map, numerically we find the tighter bound for the fidelity  $F_{\mathrm{pro}}(\mathcal{E},W(\ell))>e^{-\pi^2 \kappa/g}$ which is notably independent of $N$.

{\it Dephasing:} We next address spin decoherence. We assume that amplitude damping due to spin relaxation is small by the choice of encoding. This can be accommodated by choosing qubit states with very long decay times either as a result of selection rules, or by being far detuned from fast spin exchange transitions. Hence we will focus on dephasing.
Due to the cyclic evolution during each GPG, there is error tolerance to dephasing because if the interaction strength between the system and environment is small compared to $g$, then the spin flip pulses used between each pair of dispersive gates $R(\theta a^{\dagger}a)$ will echo out this noise to low order.  

%Consider a bath of oscillators that couple bilinearly to the spins described by $H=H_E + H^{\text{global}}_{SE}+ H^{\text{local}}_{SE}$ where the local environmental and coupling Hamiltonians are
%\begin{eqnarray}
%H_E & = & \sum_{k}\sum_{j=1}^N \omega_{j,k} b_k^{\dagger}b_k+\sum_{k} \omega_{k} c_k^{\dagger}c_k , \\
%H^{\text{global}}_{SE} & = &  J^z \sum_k   (c_k d^*_k+c^{\dagger}_k d_k), \\
%H^{\text{local}}_{SE} & = &  \sum_k \sum_{j=1}^{N} (b_{j,k} r^*_{j,k} +  b^{\dagger}_{j,k} r_{j,k}) \sigma_j^z
%\end{eqnarray}
%where $j$ is the spin index, and the local baths satisfy $[b_{j,k},b^{\dagger}_{j',k'}]=\delta_{j,j'}\delta_{k,k'}$ and the global bath $[c_{j},c^{\dagger}_{j'}]=\delta_{j,j'}$. The interaction $H^{\text{global}}_{SE}$ couples symmetrically to the spins, while $H^{\text{local}}_{SE}$ couples locally, leading to global and local dephasing respectively. 

For a given input density matrix $\rho(0)$, the output after a total time $T$ has off-diagonal matrix elements that decay as $\rho_{M,M'}(T)=\rho_{M,M'}(0)e^{-(M-M')^2A(T)}$. For the global dephasing map the numbers $M,M'\in [-N/2,N/2]$ are in the collective Dicke basis, while for local dephasing it is with respect to a local basis $M,M'\in[-1/2,1/2]$. Our argument for suppression of dephasing works for both cases. Global dephasing is the most deleterious form of noise when the state has large support over coherences in the Dicke subspace, due to decay rates that scale quadratically in the difference in $M$ number. However, it leaves the total Dicke space, and in particular the target Dicke state, invariant. Local dephasing induces coupling outside the Dicke space, but with a rate that is at most linear in $N$. 

\begin{figure}[tb]
\begin{center}
\includegraphics[width=8.6cm]{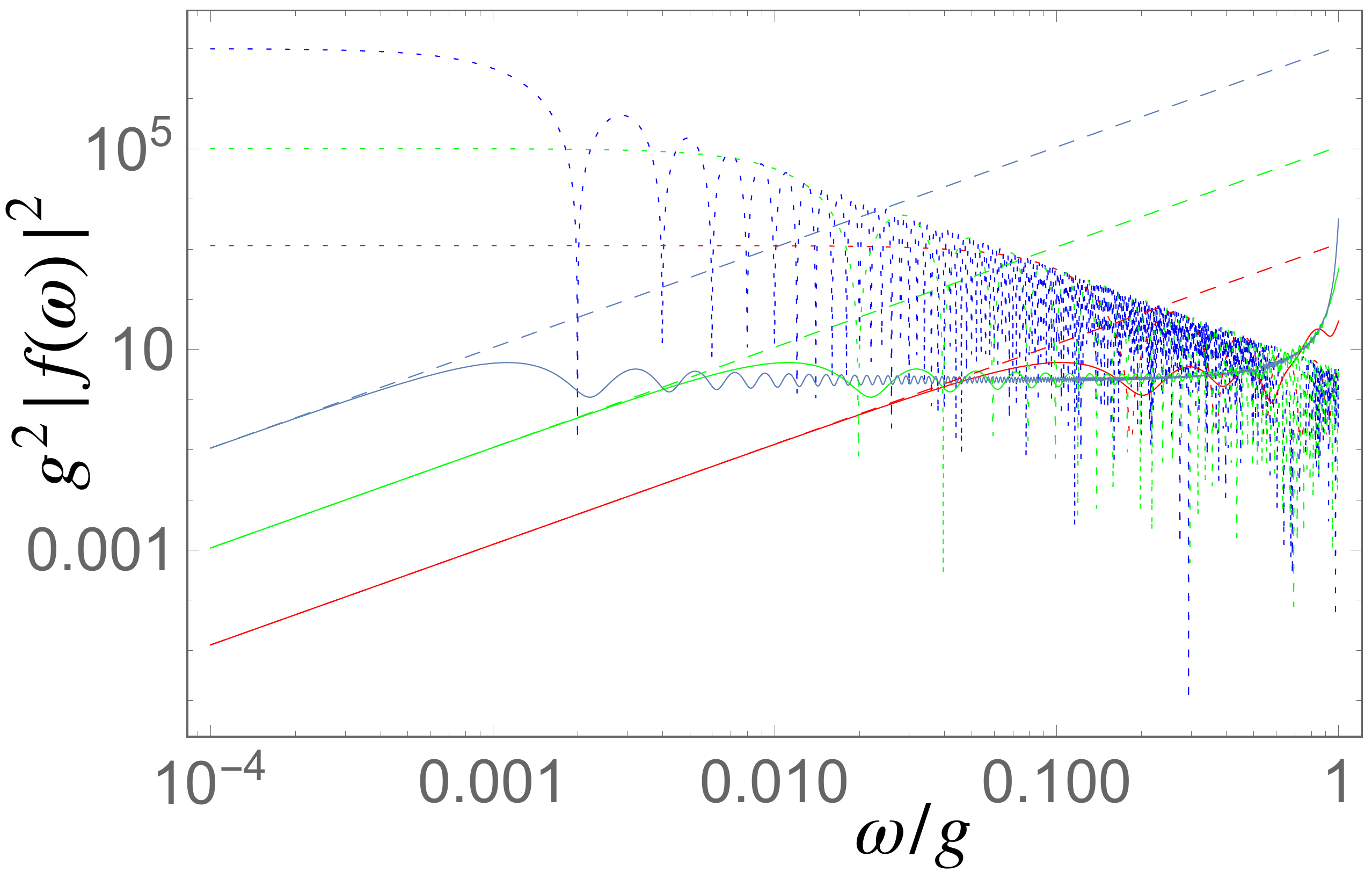}
\end{center}
\caption{Suppression of dephasing via dynamical decoupling inherent in the sequence of GPGs used for each of the operators $U_s$ and $U_w$. Solid curves are filter functions using the GPGs. Dashed curves are plots of Eq.~(\ref{filterapprox}), which is a good approximation for $\omega/g<1/\pi N$. Short-dashed curves are the bare case without decoupling. (Red, green, blue) curves correspond to $n=(10,100,1000)$ spins.}
\label{fig:5}
\end{figure}

Consider the evolution during the $N/2$ control pulses to realize either of the phasing gates $U_s$ or $U_w$.
Assuming Gaussian bath statistics, the effective dephasing rate can be written as the overlap of the noise spectrum $S(\omega)$ and the filter function $|f(\omega)|^2$ \cite{Agarwal_2010, wang_liu_2013}.
%The filter function is obtained from the windowed Fourier transform $f(w)=\int_{0}^T C(t)e^{i\omega t}$, where $C(t)$ is the time-dependent control pulse sequence.
As shown in the Supplementary material \cite{refSupplemental}, for our pulse sequence to lowest order in $\omega/g$ we find
\begin{equation}
g^2|f(\omega)|^2\approx \frac{(\omega/g)^2 \pi^4 N^2(N+2)^2}{9 (N+1)^2}.
\label{filterapprox}
\end{equation}
Comparison with the case where no spin flips are applied is plotted in Fig. \ref{fig:5} showing there is substantial decoupling from the dephasing environment when the spectral density has dominant support in the range $\omega<g/2$.

%Eq. (\reffilterapprox}) is valid for $\omega/g <1/\pi N$, and, as shown in Fig. \ref{fig:5}, for $1/\pi N<\omega/g< 1/2$ the function is essentially flat with an average value $g^2
% \overline{|f(\omega)|^2}\approx 3$ independent of $N$.  
% In the region $1/\pi N<\omega/g<1/2$ the bare filter function is oscillatory and has an average $g^2
% \overline{|f^{(0)}(\omega)|^2}\approx 13.63$, while for $\omega/g <1/\pi N$  it asymptotes to $ \frac{\pi^2 N^2(N+2)^2}{(N+1)^2}$. 
 
%In the region $\omega/g<1/\pi N$ the ratio determining the reduction factor in the dephasing rate is $\frac{|f(\omega)|^2}{|f^{(0)}(\omega)|^2}=\pi^2\omega^2/g^2$, while for $\omega/g\in[1/\pi N,1/2]$, the reduction factor can be approximated by $\frac{\overline{|f(\omega)|^2}}{\overline{|f^{(0)}(\omega)|^2}}\approx 0.22$, provided the noise spectrum is sufficiently flat there. 
The freedom to apply the GPGs in any order allows for further improvement. Consider coupling to a zero temperature Ohmic bath with noise spectrum $S(\omega)=\alpha \omega e^{-\omega/\omega_c}$ with cutoff frequency $\omega_c/g=0.1$. For $N=20$, the ratio of the effective decay rate for the linearly ordered sequence of GPGs above to that with no decoupling is $A(T)/A_0(T)=0.0085$. However, by sampling over permutations of the ordering of GPGs we find a sequence \cite{Note1}
%\footnote{The ordering $\{8,4,5,9,3,7,6,2,10,1\}$ achieves this. Note an exhaustive search over all $10!$ permutations was not done.}
achieving $A(T)/A_0(T)=0.0026$.  
Examples of the effectiveness of our decoupling protocol on sensitivity for various values of $A(T)$ are shown in Figure~\ref{fig:relative_fidelity}. Effectiveness on the fidelities can be found in the Supplementary Material \cite{refSupplemental}.

%To characterise the performance of our scheme in the presence of both mode decay $\kappa$ and effective global dephasing $A$, we performed numerical simulations of the full protocol using the joint mode-spin system with mode Fock space truncated to $15$ excitations. Results are presented in Figure~\ref{fig:relative_fidelity} and show the effectiveness of our protocol when used for metrology, and considers the uncertainty $\Delta \eta$, given a single shot measurement of ${J^z}^2$ after a collective rotation $\eta$ as defined by Eq.~(\ref{eqDeltaBetaDefinition}) on an ensemble of size $N=10$. For values of $\gamma/g \lesssim 0.01$ we beat the standard quantum limit, and for $\gamma=0$ closely approach the Cram{\'{e}}r-Rao bound. 

\begin{figure}[tb]
\begin{center}
\includegraphics[width=\columnwidth]{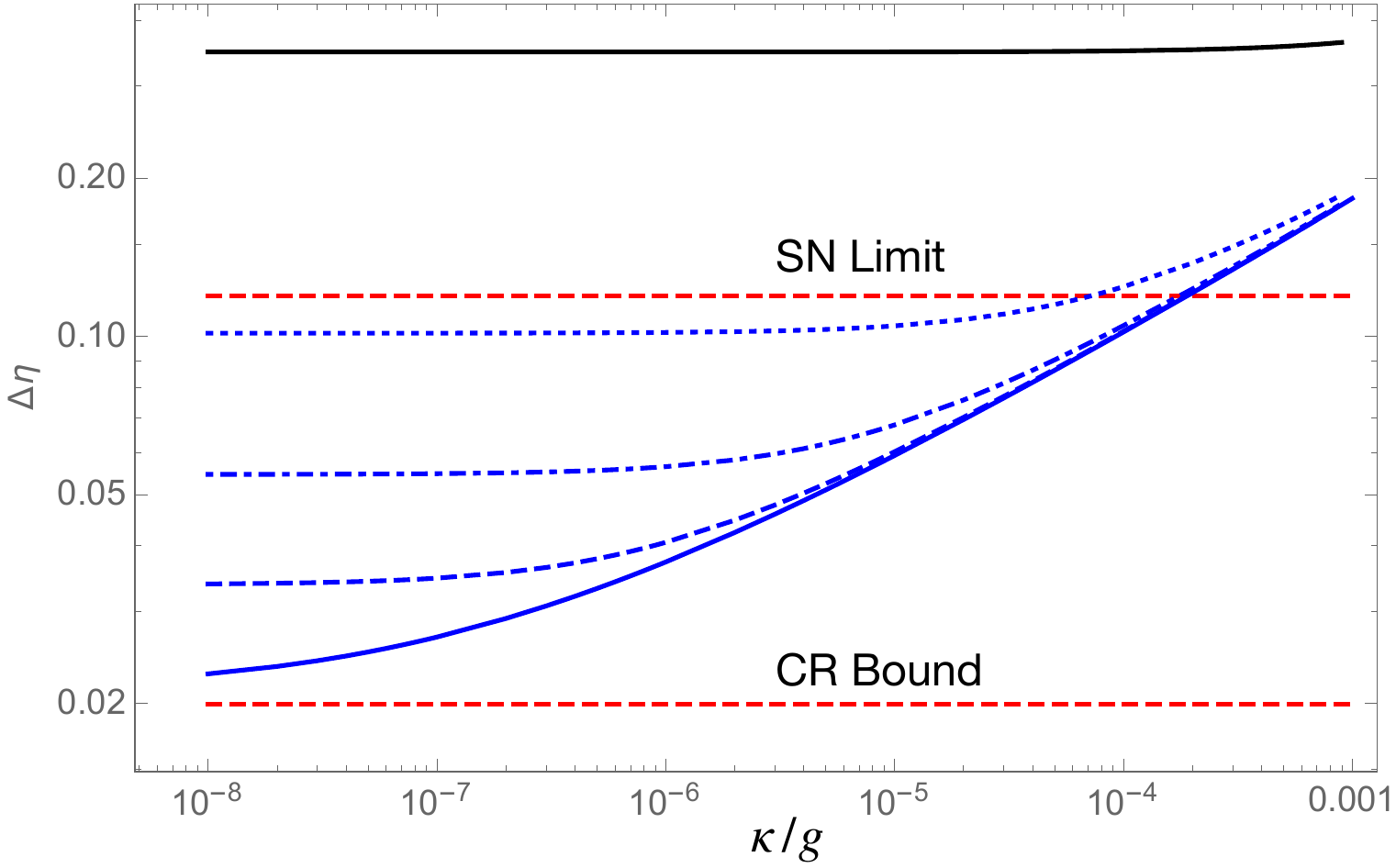}
\end{center}
\caption{Precision obtained for 70 spins with a single measurement of ${J^z}^2$ as a function of mode decay for various global dephasing factors $A(T)$: no global dephasing (blue line), $A(T)=10^{-6}$ (blue dashed), $A(T)=10^{-5}$ (blue dot-dashed), $A(T)=10^{-4}$ (blue dotted). These dephasings correspond to an underlying decoherence rate  of $\gamma_{\text{gdp}}= 10^{-4}g$ accumulated over each phasing gate of duration $T$. For a zero temperature Ohmic bath, the corresponding cuttoff frequencies are: $\omega_c/g=\{0.003,0.022,0.094\}$. Black line: performance without dynamical decoupling with $A_0(T)=0.0223$.}
%e.g. if one were to switch the sign of the dispersive coupling during each GPG rather than flipping the spins. 
%(b) Fidelity error for the same environments as above.
\label{fig:relative_fidelity}
\end{figure}

{\it {Error-tolerant states}}.--- The state preparation method we have described has inherent tolerance to decoherence. However once the state is prepared further errors such as qubit loss or dephasing could accumulate.
%, while waiting for the accumulation of the measurement signal. 
Strategies to address this by using superpositions of Dicke states were recently proposed \cite{ouyang2019}. The states considered were
\[
\ket{\varphi_u}=\frac{1}{\sqrt{2^n}}\sum_{j=0}^n\sqrt{\binom{n}{j}}\ket{J=\frac{knu}{2},M=kj-J}.
\]
The number of spins $N=k\times n\times u$, and the parameters $u$ and $n$ determine the robustness of the states to some number of loss and dephasing errors respectively.
%, while $k$ is a parameter to scale the number of qubits in the superposition.
%The case $u=1$ tolerates erasure errors; specifically, the state $\ket{\varphi_1}$ has a large quantum Fisher information obeying Heisenberg scaling when the number of erasure errors is less than $n$. 
A state performing well in the presence of one erasure error is  $u=1, n=2, k=N/2$, which can be written $\ket{\varphi_1}=\frac{1}{2}(\ket{J,-J}+\sqrt{2}\ket{J,0}+\ket{J,J})$
while $u=2,n=1,k=N/2$ tolerates one dephasing error and can be written $\ket{\varphi_2}=\frac{1}{\sqrt{2}}(\ket{J,-J}+\ket{J,0})$.
Both of these states can be prepared using our protocol. The specific steps are given in the Supplementary Material \cite{refSupplemental}. Superpositions of Dicke states also arise as code words of so-called permutationally invariant quantum codes \cite{Ruskai}, and of recently discovered codes which admit Gaussian Clifford operations \cite{Gross} which our method could also prepare.

{\it {Implementations}}.--- The scheme we have presented is amenable to a variety of architectures which allow collective dispersive couplings between spins and an oscillator. These include: trapped Rydberg atoms coupled to an microwave cavity \cite{Sayrin:2011eu, garcia2019}, trapped ions coupled to a common motional mode \cite{Pedernales:2015si} or to an optical cavity mode \cite{PhysRevLett.122.153603}, superconducting qubits coupled to microwave resonators \cite{Wang1087}, and NV centres in diamond coupled to a microwave mode inside a superconducting transmission line cavity \cite{PhysRevLett.118.140502}.

One contender to test our scheme is Rydberg atoms coupled to microwave cavities. Recently the dispersive detection of small atomic Rydberg ensembles coupled to a high-Q microwave cavity
%(Q-factor $1.7 \times 10^6$) 
was reported \cite{garcia2019}. Their numbers suggest a ratio of
%cavity decay rate to single-atom dispersive coupling strength of
$\kappa/g \approx 0.8$ (with $\kappa = 2\pi \times 11.8$~kHz and $g = 2\pi \times 14.3$~kHz). The collective coupling rate observed was a few MHz, suggesting an additional pathway to improving $\gamma/g$ by orders of magnitude by encoding spins through collective subensembles.   
%Since the performance improves with stronger dispersive coupling strength $g$, it can be %advantageous to choose spin encodings which enhance the interaction. 
Consider an encoding where each spin is composed of $n$ physical spins with logical states $\ket{0}=\ket{j=n/2,-j}$ and $\ket{1}=\ket{j=n/2,-j+1}$, i.e. the permutationally invariant states of zero or one excitation shared among the $n$ spins. If the spins within each logical qubit interact, e.g. via dipole-dipole interactions, there will be a dipole-blockade to larger numbers of excitations. Hence collective rotations frequency-tuned to the transition energy $E_1-E_0$ will be collective but only act on this qubit subspace. The dispersive interaction strength is enhanced by $g\rightarrow g\sqrt{n}$, provided
%Since we have assumed that the rotations and  dispersive couplings are equivalent on all logical spins, it will be important
the number $n$ is similar for all logical spins. Using this kind of encoding, dispersive coupling with strength $g\approx 2\pi\times 2.2$\ MHz was obtained with NV ensembles in diamond bonded onto a transmission line resonator with quality factor $Q\approx 4300$ at the first harmonic frequency $\omega_1=2\pi\times 2.75$\ GHz. Microwave cavities with much higher quality factors, e.g. $Q\approx 10^9$, have been realized \cite{Castelli} which for the same dispersive coupling could provide $\kappa/g\approx 10^{-6}$.

\begin{acknowledgments}
We acknowledge helpful discussions with Jason Twamley and Yingkai Ouyang. This research was funded in part by the Australian Research Council Centre of Excellence for Engineered Quantum Systems (Project number CE170100009).
\end{acknowledgments}

%\section{Data availability}
%The datasets generated during and/or analyzed during the current study are available from the corresponding author on reasonable request.

%\bibliographystyle{apsrev4-1}
%\bibliography{Superabsorb}

%\section{Author contributions} 
%GKB and MTJ did analytic modelling and also carried out the numerical simulations. All authors contributed to the theoretical development of this work. 

%\section{Competing interests} 
%The authors declare no competing interests.

\section{Supplemental Information}
\subsection{Gate fidelity with cavity decay}\label{Sec:CavDec}

Cavity field decay at a rate $\kappa$ acts as a source of error for the many body
interactions which the cavity mode mediates.
%For the protocols above where the system of spins interact
% with the cavity field, the joint state can be decomposed as
% \begin{multline}
% \rho(t)  = \sum_{M,M^{\prime
% }}\rho_{M,M^{\prime }}(0)\Big(%
% \ket{J,M}\bra{J,M'}\otimes \ket{\alpha_{M}}\bra{\beta_{M'}}\Big)(t).%
% \label{rho}
% \end{multline}
% The field states $\ket{\alpha_{M}}$, $\ket{\beta_{M}}$, may depend on the angular
% momentum projections. We describe evolution for the case where the field
% states are Fock states for the single photon probe protocol and when they
% are coherent states for the geometric phase gate. Depending on the protocol
% used, the field states themselves may be entangled with the state of an
% ancillary spin but we focus on computing fidelities for evolution steps
% where the ancilla is non interacting.
Consider the the joint evolution of the spins and the mode. The coupling of the mode to its environment is treated as irreversible and thus can be described by the standard master equation in Lindblad form.  The equation of
motion for the joint state is
\begin{align}
\dot{\rho}(t) & = \mathcal{L}(\rho(t)) \nonumber\\
& =  -i[V,\rho(t)]+\frac{\kappa}{2} (2a \rho(t) a^{\dagger}-a^{\dagger} a
\rho(t)-\rho(t)a^{\dagger}a).\label{rhodot}
\end{align}
The evolution due to decay conserves the quantum number $J$, and it will be convenient to compute the adjoint action on on a joint state state of the spins and mode with Heisenberg evolved operators $e^{\mathcal{L}t}A^{M,M^{\prime }}(0)$ where:
\[
A^{M,M^{\prime }}(t)\equiv \ket{J,M}\bra{J,M'}\otimes \ket{\alpha_{M}}\bra{\beta_{M'}}(t).
\]
The solutions are easily verified to be given by
\begin{multline}
A^{M,M^{\prime }}(t) = \sum_{n=0}^\infty\frac{b^n_{MM^{\prime
}}(t)}{n!} e^{-(ig M+\kappa/2)a^\dagger a t} \\
\times a^nA^{M,M^{\prime }}(0){(a^\dagger)}^ne^{(i g M^{\prime
}-\kappa/2) a^\dagger a t}
\label{evolveA}
\end{multline}
where
\begin{equation}
b_{M M^{\prime }}(t)=\frac{\kappa(1-e^{-[\kappa+ig(M-M^{\prime
})]t})}{\kappa+ig(M-M^{\prime })}.
\end{equation}
The evolved state is then
\[
\rho(t)=e^{\mathcal{L}t}\rho(0)=\sum_{M,,M^{\prime }}\rho_{M,M^{\prime }}(0)A^{M,M^{\prime }}(t).
\]

\label{gpfid}

In order to evaluate the effect of cavity decay during the the geometric
phase gate, we are particularly interested in the case where initially $%
A^{M,M^{\prime }}(0)=\ket{M}\bra{M'}%
\otimes|\alpha_{M}\rangle\langle\beta_{M^{\prime }}|$, with $%
|\alpha_{M}\rangle$, $\ket{\beta_{M'}}$ coherent states. This kind of
factorization is true at any stage of spin coupling to the field. Using Eq.~(\ref{evolveA}), the sum becomes an exponential and the evolved state is
\begin{align}
\rho(t) & = e^{\mathcal{L}t}\rho(0) \nonumber\\
& = \sum_{M,M^{\prime
}}e^{d_{M,M'}(t)}\rho_{M,M^{\prime
}}(0)\ket{J,M}\bra{J,M'} \nonumber\\
& \hspace{0.5in}  \otimes |e^{-(igM+\kappa/2)t}\alpha_{M}\rangle\langle
e^{-(igM^{\prime }+\kappa/2)t}\beta_{M^{\prime }}|,%
\label{evolved}
\end{align}
where
\begin{equation}
d_{M,M'}(t)=\alpha_{M}\beta_{M^{\prime }}^*b_{M,M^{\prime }}(t)
-(|\alpha_{M}|^2+|\beta_{M^{\prime }}|^2){\textstyle\frac{1-e^{-\kappa t}%
}{2}}.
\end{equation}

We ignore decay during the displacement stages of the evolution (i.e. we
assume these are done quickly relative to the decay rate), and we assume
that the system particles do not interact with the field during these steps.
For simplicity we evaluate the performance when the cavity begins in the vacuum state, in which case there are seven time steps to consider:
\[
D(-\beta)e^{-i \tau_5 V}D(-\alpha)e^{i \tau_3 V}D(\beta)e^{-i \tau_1
V}D(\alpha).
\]
 Let $\tau_5=\tau_3=\tau_1$ so that the
periods of spin field coupling are all equal in duration. In order that the field state return to
the vacuum at the end of the sequence, we choose $\alpha^{\prime}=\alpha^{ -\kappa
\tau_1},\beta^{\prime}=\beta^{ -\kappa \tau_1}$ for the parameters of the second
two displacement operators. The total sequence then yields the
output state:

\begin{equation}
\begin{array}{lll}
\rho_{\mathrm{out}}  &=&  \displaystyle{\sum_{M,M^{\prime }}} \rho_{M,M^{\prime }}(0)R_{M,M^{\prime }}\ket{J,M}%
\bra{J,M'}\otimes \ket{{\rm vac}}\bra{{\rm vac}} \\
&&\times e^{-i2\chi(\sin(\phi+g\tau_1 M)-\sin(\phi+g\tau_1 M^{\prime }))},%
\end{array}
\end{equation}

where we defined $R_{M,M^{\prime }}=e^{d_{M,M'}(t_2)+d_{M,M'}(t_4)+d_{M,M'}(t_6)}$ and $%
\chi=|\alpha \beta |(e^{-3\kappa \tau_1/2}+e^{-\kappa \tau_1/2})/2$. This can be interpreted as coherent evolution %by the Hamiltonian
with an evolution operator
\[
U=e^{-i2\chi\sin(\phi+\theta J^z)},
\]
where $\theta=g\tau_1$, followed by further evolution diagonal in the $%
\{M\}$ basis and dephasing. Matrix elements diagonal in $M$ are invariant.

For simplicity, we assume $|\alpha|=|\beta|$, $g>0$ and write $\theta=g\tau_1$. The factor $R_{M,M'}$ that dictates the deviation from perfect evolution can be written
\begin{equation}
R_{M,M'}=e^{-\Gamma_{M,M'}}e^{i\Delta_{M,M'}}
\end{equation}
where $\Gamma_{M,M'}$ and $\Delta_{M,M'}$ are real, and explicitly are
\begin{widetext}
\begin{equation}
\begin{array}{lll}
\Gamma_{M,M'}&=&\frac{|\alpha|^2 (M-M') e^{-i (\phi +\theta  (M'+M-2 i \frac{\kappa}{g}))}}{2 ((M-M')^2+(\frac{\kappa}{g})^2)}\\
&&\Big(-4 (M-M') e^{i (\theta  (M'+M)+\phi )}-4 i \frac{\kappa}{g} e^{2 i \theta  M'+\theta  \frac{\kappa}{g}+i \phi }+(M'-M+i \frac{\kappa}{g}) e^{\frac{1}{2} \theta  (2 i M'+4 i M+\frac{\kappa}{g})+2 i \phi }\\
&&+(-M'+M+i \frac{\kappa}{g}) e^{i \theta  M'+2 i \theta  M+\frac{3 \theta  \frac{\kappa}{g}}{2}+2 i \phi }+(M'-M-i \frac{\kappa}{g}) e^{\frac{1}{2} \theta  (4 i M'+2 i M+\frac{\kappa}{g})+2 i \phi }\\
&&+(-M'+M-i \frac{\kappa}{g}) e^{2 i \theta  M'+i \theta  M+\frac{3 \theta  \frac{\kappa}{g}}{2}+2 i \phi }+4 (M-M') e^{i (\phi +\theta  (M'+M-2 i \frac{\kappa}{g}))}+e^{\frac{1}{2} \theta  (\frac{\kappa}{g}+2 i M)} (M'-M+i \frac{\kappa}{g})\\
&&+e^{\frac{3 \theta  \frac{\kappa}{g}}{2}+i \theta  M} (-M'+M+i \frac{\kappa}{g})+(M'-M-i \frac{\kappa}{g}) e^{\frac{1}{2} \theta  (\frac{\kappa}{g}+2 i M')}+(-M'+M-i \frac{\kappa}{g}) e^{\frac{3 \theta  \frac{\kappa}{g}}{2}+i \theta  M'}\\
&&+4 i \frac{\kappa}{g} e^{2 i \theta  M+\theta  \frac{\kappa}{g}+i \phi }\Big)
\end{array}
\end{equation}

\begin{equation}
\begin{array}{lll}
\Delta_{M,M'}&=&-\frac{|\alpha|^2(1+e^{i (\theta  (M'+M)+2 \phi )})e^{-i \theta  (M'+M)-\frac{3 \theta  \frac{\kappa}{g}}{2}-i \phi }}{2 ((M-M')^2+(\frac{\kappa}{g})^2)}\\
&&\Big(e^{\theta  (\frac{\kappa}{g}+i M)} (2 M^2-(4 M+i \frac{\kappa}{g}) M'+2 (M')^2+i M \frac{\kappa}{g}+(\frac{\kappa}{g})^2)-(2 M^2+(-4 M+i \frac{\kappa}{g}) M'+2 (M')^2-i M \frac{\kappa}{g}+(\frac{\kappa}{g})^2)\\
&&e^{\theta  (\frac{\kappa}{g}+i M')}-e^{i \theta  M} (2 M^2-(4 M+i \frac{\kappa}{g}) M'+2 (M')^2+i M \frac{\kappa}{g}+3 (\frac{\kappa}{g})^2)\\
&&+e^{i \theta  M'} (2 M^2+(-4 M+i \frac{\kappa}{g}) M'+2 (M')^2-i M \frac{\kappa}{g}+3 (\frac{\kappa}{g})^2)\Big)
\end{array}
\end{equation}
Notice, $\Gamma_{M,M'}=\Gamma_{M',M}$ and $\Gamma_{M,M}=0$ and also $\Delta_{M,M'}=-\Delta_{M',M}$. An expansion up to first order in  $\frac{\kappa}{g}$ yields simplified expressions:
\begin{equation}
\begin{array}{lll}
\Gamma_{M,M'}&=&
\frac{|\alpha|^2 \frac{\kappa}{g}}{M-M'}
\Big(2 \sin  (\theta  M'+\phi )-\theta  M' (\cos  (\theta  M'+\phi )+\cos  (\theta  M+\phi )+4)+\theta  M (\cos  (\theta  M'+\phi ))\\
&&-4 (\sin  (\theta  (M-M')))-2 (\sin  (\theta  M+\phi ))+\theta  M (\cos  (\theta  M+\phi ))+4 \theta  M\Big)
\end{array}
\end{equation}
\begin{equation}
\begin{array}{lll}
\Delta_{M,M'}&=&
|\alpha|^2 \theta  \frac{\kappa}{g} (-\sin  (\theta  M')+i (\cos  (\theta  M'))+\sin  (\theta  M)-i (\cos  (\theta  M))) (\cos  (\theta  M'+\theta  M+\phi )\\
&&-i (\sin  (\theta  M'+\theta  M+\phi ))) (i (\sin  (\theta  (M'+M)+2 \phi ))+\cos  (\theta  (M'+M)+2 \phi )+1)
\end{array}
\end{equation}
\end{widetext}

Now, one can check that the decoherence factors are bounded as follows: $\Gamma_{M,M'}\leq |\alpha|^2 6\pi \kappa/g $
and $|\Delta_{M,M'}|\leq |\alpha|^2 4\pi \kappa/g $. 

In order to obtain a more rigorous process fidelity bound, we proceed as follows.
For an input spin state in the symmetric Dicke space $\rho=\sum_{M,M'}\rho_{M,M'}\ket{J,M}\bra{J,M'}$, the process for the $k-$th GPG with decay on the spins, including the modified displacement operations above, is \cite{Brennen:09}
\begin{equation}
\begin{array}{lll}
\mathcal{E}^{(k)}(\rho)&=&U_{GPG}(\theta_k,\phi_k,\chi_k) \big[ \sum_{M,M'}R^{(k)}_{M,M'}\rho_{M,M'}.  \\
&&\ket{J,M}\bra{J,M'} \big] \times U_{GPG}^{\dagger}(\theta_k,\phi_k,\chi_k)
\end{array}
\end{equation}
%We achieve best performance by choosing the displacement amplitudes according to
%\begin{equation}
%|\alpha_k\beta_k|=\frac{2\pi}{(n+1)(e^{-3\Gamma_{\rm gd}\theta_k/2g}+e^{-\Gamma_{\rm gd}\theta_k/2g})}.
%\end{equation}
where
\[
\chi_k=|\alpha_k|^2(e^{-3\theta_k\kappa/2g}+e^{-\theta_k\kappa/2g})/2
\]
and 
\[
R^{(k)}_{M,M'}=e^{-\Gamma_{M,M'}(\theta_k,\phi_k,\alpha_k)}e^{i\Delta_{M,M'}(\theta_k,\phi_k,\alpha_k)}
\]
with the factors $\Gamma_{M,M'}$ and $\Delta_{M,M'}$ given above.
%For $M\neq M'$ we find for $\kappa/g\ll 1$: 
%\begin{equation}
%\begin{array}{lll}
%\Gamma_{M,M'}&=&
%\frac{|\alpha|^2 \frac{\kappa}{g}}{M-M'}
%\Big(2 \sin  (\theta  M'+\phi )-\theta  M' (\cos  (\theta  M'+\phi ) \\
%&&+\cos  (\theta  M+\phi )+4)+\theta  M \cos  (\theta  M'+\phi ) \\
%&& -4 \sin  (\theta  (M-M')) - 2 \sin  (\theta  M+\phi) \\
%&&+\theta  M (\cos  (\theta  M+\phi ))+4 \theta  M\Big)
%\end{array}
%\end{equation}
%\begin{equation}
%\begin{array}{lll}
%\Delta_{M,M'}&=&
%|\alpha|^2 \theta  \frac{\kappa}{g} (-\sin  (\theta  M')+i \cos  (\theta  M') \\
%&&+\sin  (\theta  M)-i \cos  (\theta  M)) (\cos  (\theta  M'+\theta  M+\phi )\\
%&&-i \sin  (\theta  M'+\theta  M+\phi )) (i \sin  (\theta  (M'+M)\\
%&&+2 \phi ) + \cos  (\theta  (M'+M)+2 \phi ) + 1).
%\end{array}
%\end{equation}
Now if we adjust $\alpha$ such that on the $k$-th GPG, $\chi_k=\pi/(N+1)$, then we have
\[
|\alpha_k|^2=\frac{2\pi}{(N+1)(e^{-3\theta_k\kappa/2g}+e^{-\theta_k\kappa/2g})}.
\]
Because the coherent and decoherent maps for different GPGs commute, the entire sequence that phases a Dicke state according to $W(\ell)$ is:
\begin{equation}
\begin{array}{lll}
\mathcal{E}(\rho)&=&\mathcal{E}^{(N/2)}\circ\cdots\circ \mathcal{E}^{(1)}(\rho)\\
&=&W(\ell) \sum_{M,M'}\Upsilon_{M,M'}(\ell)\rho_{M,M'}\ket{J,M}\bra{J,M'} W(\ell)^{\dagger}
\end{array}
\end{equation}
This describes ideal evolution followed by a nonlinear dephasing map, where the decoherence factor is
\begin{equation}
    \begin{array}{lll}
    \Upsilon_{M,M'}(\ell)&=&\prod_{k=1}^{N/2}R^{(k)}_{M,M'}\\
    &=& \exp \bigl[ \sum_{k=1}^{N/2}(-\Gamma_{M,M'}(\theta_k,\phi_k(\ell),\alpha_k)  \nonumber \\
    && \hspace{1cm} + i\Delta_{M,M'}(\theta_k,\phi_k(\ell),\alpha_k)) \bigr].
    \end{array}
\end{equation}

The process fidelity $F_{\mathrm{pro}}(\mathcal{E},U)$ measures how close a
quantum operation $\mathcal{E}$ is to the ideal operation $U$ as measured by
some suitable metric. The fidelity measure we use is the overlap between the
induced Jamio\l kowski-Choi state representations of the operations.
The process fidelity is readily computed using the fact that the noise map $%
\mathcal{E}(\rho_S(0))$ commutes with the target unitary $U$. Hence, we can
compute the fidelity which measures how close the noisy map $\mathcal{E}%
^{\prime }(\rho_S(0))=U^{\dagger} \mathcal{E}(\rho_S(0)) U$ is to the ideal
operation, i.e. the identity operation:
\[
F_{\mathrm{pro}}(\mathcal{E},U)=F_{\mathrm{pro}}(\mathcal{E}^{\prime },%
\mathcal{I})=_{S,S^{\prime }}\langle \Phi^{+}|\rho_{\mathcal{E}^{\prime
}}|\Phi^{+}\rangle_{S,S^{\prime }}.
\]
where
\begin{equation}
    |\Phi^+\rangle_{S,S^{\prime }}  =  \frac{1}{\sqrt{D}}\sum_{M}%
\ket{J,M}_S\otimes \ket{J,M}_{S^{\prime }},
\end{equation}
Here we are computing the overlap of the Jamio\l kowski-Choi representations
of the maps as states in the Hilbert space $\mathcal{H}_S\otimes \mathcal{H}%
_{S^{\prime }}$ containing our system space and a copy each with dimension $%
D $:
\begin{align*}
\rho_{\mathcal{E}^{\prime }}  &=  \mathcal{I}_{S}\otimes\mathcal{E}^{\prime}%
_{S^{\prime }}(\ket{\Phi^+}_{S,S^{\prime }}) \\
 &=  \frac{1}{D}\sum_{M,M^{\prime
}}\Upsilon_{M,M^{\prime }}(\ell)\phantom{=}\ket{M}_S \bra{M'}\otimes \ket{M}_{S^{\prime }} \bra{M'}.%
\end{align*}%
Hence
\[
F_{\mathrm{pro}}(\mathcal{E},W(\ell))=\frac{1}{(N+1)^2}\sum_{M,M'=-J}^J \Upsilon_{M,M^{\prime }}(\ell),
\]
From the limits above, we readily find the process fidelity bound
\begin{equation}
F_{\mathrm{pro}}(\mathcal{E},U_{\rm GPG})>e^{-6\pi|\alpha|^2 \kappa/g}\cos(|\alpha|^24\pi \kappa/g).
\label{fullfid}
\end{equation}
Finally we note that numerical simluation suggests a tighter bound given by 
\begin{equation}
F_{\mathrm{pro}}(\mathcal{E},W(\ell))>e^{-\pi^2 \kappa/g}.
\end{equation}
Notably, this fidelity is independent of $N$.

\subsection{A loose upper bound on precision as function of fidelity}\label{Sec:fidelity_estimate}

The precision of the estimation parameter $\eta$ is expressed as
\begin{align}
	\nonumber (\Delta \eta)^2=&\big((\Delta {J^x}^2)^2 f(\eta)+4\langle {J^x}^2\rangle-3\langle {J^y}^2\rangle-2\langle {J^z}^2\rangle  \\
	&\times(1+\langle {J^x}^2\rangle)+6\langle J^z{J^x}^2J^z\rangle\big)(4(\langle {J^x}^2\rangle-\langle {J^z}^2\rangle)^2)^{-1}
	\label{eq:precision_sq}
\end{align}

To check how the precision scales in relation to the fidelity, $F$, we can calculate the precision assuming an input density matrix, $\rho=a \ket{J,0}\bra{J,0} + b \mathbb{1}$, where $\ket{J,0}$ is our ideal Dicke state, $\mathbb{1}$ is the identity matrix of size $(N+1) \times (N+1)$ and $a$ and $b$ are related to fidelity as 
$a=(1+1/N) F- 1/N$ and $b=(1-F)/N$.
We choose this form for the input density matrix so that applying a global dephasing map to the output state of our protocol would make it diagonal in the $\ket{J,M}$ basis but keep the population in $\ket{J,0}$ constant. Assuming the diagonal matrix elements (except the $\ket{J,0}\bra{J,0}$ entry) are equally weighted is a maximally unbiased assumption. After calculating the variances and expectation values of the angular momentum operators as they appear in Equation~(\ref{eq:precision_sq}), then taking the high fidelity limit $F\rightarrow 1$ and assuming large $N$,  the precision is found to be
\begin{equation}
(\Delta \eta)^2=2/(N (N+2))  + \sqrt{(1-F)/10},
\label{eq:approximate_prec}
\end{equation}
where $2/(N (N+2))$ is the Cram\'er-Rao bound. Numerically we find this approximate form is extremely good for $1-F<10^{-2}$.

% In Figure \ref{fig:prec_plots} we plot precision $(\Delta \eta)^2$ versus number of spins $N$ for two different fidelities. The top set of curves correspond to $F=0.99$ and the bottom set correspond to $F=0.999$. There are three curves for each value of $F$,  with the blue curve showing the exact expression for $(\Delta \eta)^2$, the orange curve showing the approximate expression for $(\Delta \eta)^2$ after doing a Taylor series expansion in the parameter $\epsilon=1-F$ and only keeping terms up to order $\sqrt{\epsilon}$, and the green curve represents the expression for $(\Delta \eta)^2$ approximated for large $N$ as it appears in Eq.~(\ref{eq:approximate_prec}). As can be seen, for both fidelities, the approximate curves are a very good fit to the exact curves (in the $F=0.999$ case they lie essentially on top of each other).
For our choice of the input density matrix, there appears to be a lower bound to the precision (as a function of $N$) that is set by the fidelity.
If we would want the overall expression to fall off as $1/N^2$, i.e to achieve the Heisenberg scaling, then we would need the error $1-F$ to scale as $1/N^4$. While this requirement is demanding in terms of performance, it should be noted that we have assumed that all the populations in $\ket{J,M},\ M\neq 0$ are equal when in fact the non-zero terms in the
output density matrix are much more concentrated
near the target state $\ket{J,0}$ for our protocol. As the precision involves 
terms like expectation values of $J^{z^4}$,  error terms with support on states far away from $\ket{J,0}$ will give large
errors. Thus, we are overestimating the error in this case and 
this should be viewed as a loose upper bound on the precision.

\subsection{Dephasing rates and compensation via dynamical decoupling}
\label{Sec:dynamical_decoupling}

Due to the cyclic evolution during each GPG, there is error tolerance to dephasing because if the interaction strength between the system and environment is small compared to $g$, then the spin flip pulses used between each pair of dispersive gates $R(\theta a^{\dagger}a)$ will echo out this noise to low order.  

%Consider a bath of oscillators that couple bilinearly to the spins described by $H=H_E + H^{\text{global}}_{SE}+ H^{\text{local}}_{SE}$ where the local environmental and coupling Hamiltonians are
%\begin{eqnarray}
%H_E & = & \sum_{k}\sum_{j=1}^N \omega_{j,k} b_k^{\dagger}b_k+\sum_{k} \omega_{k} c_k^{\dagger}c_k , \\
%H^{\text{global}}_{SE} & = &  J^z \sum_k   (c_k d^*_k+c^{\dagger}_k d_k), \\
%H^{\text{local}}_{SE} & = &  \sum_k \sum_{j=1}^{N} (b_{j,k} r^*_{j,k} +  b^{\dagger}_{j,k} r_{j,k}) \sigma_j^z
%\end{eqnarray}
%where $j$ is the spin index, and the local baths satisfy $[b_{j,k},b^{\dagger}_{j',k'}]=\delta_{j,j'}\delta_{k,k'}$ and the global bath $[c_{j},c^{\dagger}_{j'}]=\delta_{j,j'}$. The interaction $H^{\text{global}}_{SE}$ couples symmetrically to the spins, while $H^{\text{local}}_{SE}$ couples locally, leading to global and local dephasing respectively. 

 For a given input density matrix $\rho(0)$, the output after a total time $T$ has off-diagonal matrix elements that decay as $\rho_{M,M'}(T)=\rho_{M,M'}(0)e^{-(M-M')^2A(T)}$. For the global dephasing map the numbers $M,M'\in [-N/2,N/2]$ are in the collective Dicke basis, while for local dephasing it is with respect to a local basis $M,M'\in[-1/2,1/2]$. Our argument for suppression of dephasing works for both cases. Global dephasing is the most deleterious form of noise when the state has large support over coherences in the Dicke subspace, due to decay rates that scale quadratically in the difference in $M$ number. However, it leaves the total Dicke space, and in particular the target Dicke state, invariant. In contrast, local dephasing induces coupling outside the Dicke space, but with a rate that is at most linear in $N$. 

Consider the evolution during the $N/2$ control pulses to realize either of the phasing gates $U_s$ or $U_w$.
Assuming Guassian bath statistics, the effective dephasing rate can be written as the overlap of the noise spectrum $S(\omega)$ and the filter function $|f(\omega)|^2$ (see e.g. \cite{Agarwal_2010, wang_liu_2013}):
\begin{equation}
    A(T)=\frac{1}{2\pi}\int_0^{\infty} d\omega S(\omega)|f(\omega)|^2.
\end{equation}
For an initial system-bath state $\rho(0)=\rho_S(0)\otimes\rho_B(0)$ with the bath in thermal equilibrium $\rho_B(0)=\prod_{k} (1-e^{-\beta\omega_k})e^{-\beta \omega_k b^{\dagger}_kb_k}$ at inverse temperature $\beta$ ($k_B\equiv 1$), the noise spectrum is $S(w)=2\pi (n(\omega)+1/2)I(\omega),$ where $I(\omega)=\sum_k |g_k|^2\delta(\omega-\omega_k)$ is the boson spectral density, and $n(\omega_k)=(e^{\beta \omega_k}-1)^{-1}$ is the thermal occupation number in bath mode $k$. The filter function is obtained from the windowed Fourier transform $f(w)=\int_{0}^T C(t)e^{i\omega t}$, where $C(t)$ is the time-dependent control pulse sequence. In the present case $C(t)$ is a unit sign function that flips every time a collective spin flip is applied:
 \[
 C(t)=\left\{\begin{array}{cc}1 & t\in \cup_{k=1}^{N/2} \{[T^{(0)}_k,T^{(1)}_k)\cup [T^{(2)}_k,T^{(3)}_k)\} \\-1 & t\in\cup_{k=1}^{N/2}\{[T^{(1)}_k,T^{(2)}_k)\cup [T^{(3)}_k,T^{(4)}_k)\} \\0 & {\rm otherwise}\end{array}\right.
 \]
 where $T^{(m)}_k=m \theta_k/g+4\sum_{j=1}^{k-1} \theta_j/g$ are the flip times with the duration between pulses growing linearly. The angles are $\theta_k=\frac{2\pi k}{N+1}$ and the total time is
 \[
 T=T^{(4)}_{N/2}=\frac{\pi N(N+2)}{g (N+1)}.
 \]
 The explicit form of the filter function is
 \[
 \begin{array}{lll}
|f(\omega)|^2&=&\frac{1}{\omega^2}\Big|\sum_{k=1}^{N/2}(e^{i\omega T^{(0)}_k}-2e^{i\omega T^{(1)}_k}+2 e^{i\omega T^{(2)}_k}\\
&&-2 e^{i\omega T^{(3)}_k}+e^{i\omega T^{(4)}_k})\Big|^2.
\end{array}
\]

\begin{figure}[tb]
\begin{center}
\includegraphics[width=\columnwidth]{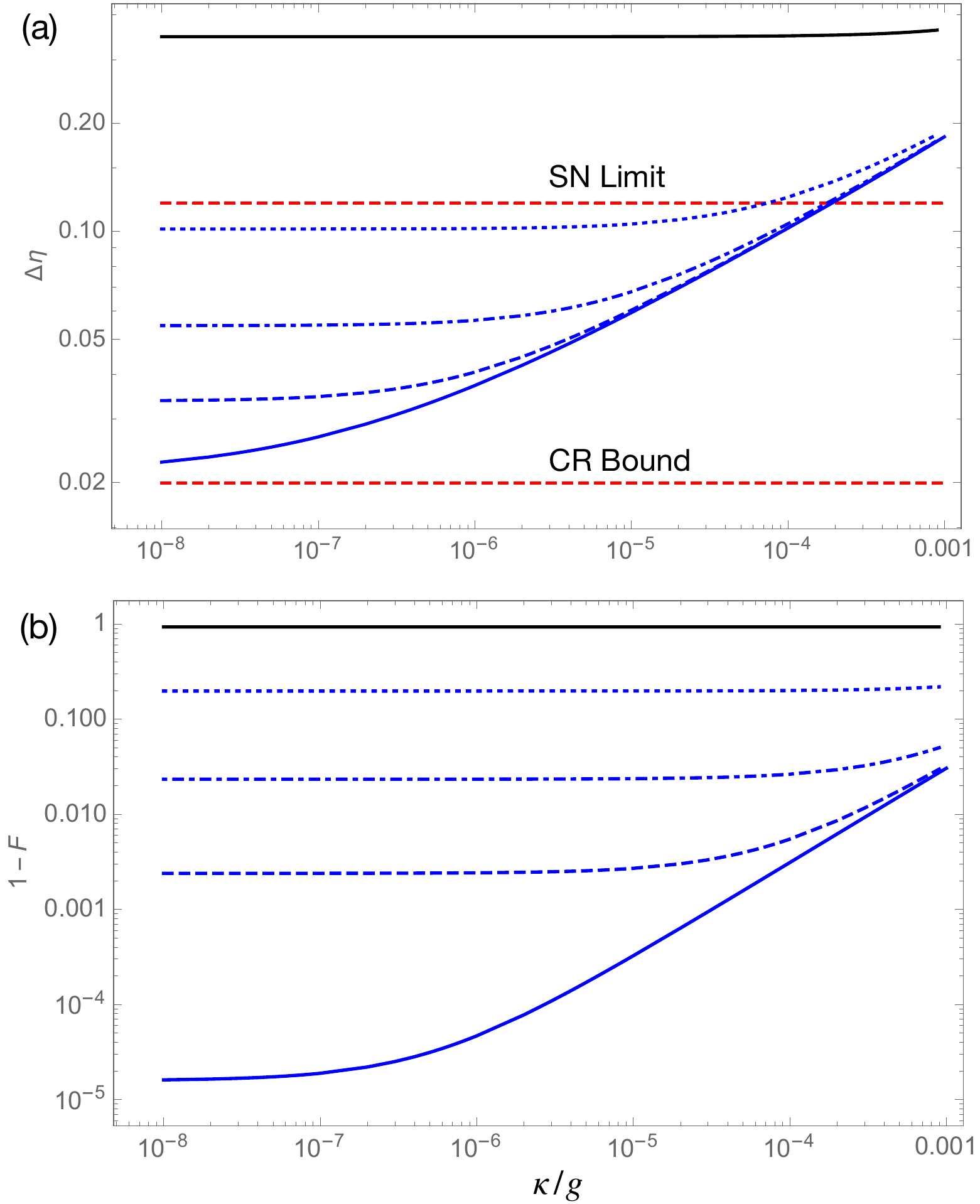}
\end{center}
\caption{a) Precision obtained for 70 spins with a single measurement of ${J^z}^2$ as a function of mode decay for various global dephasing factors $A(T)$: no global dephasing (blue line), $A(T)=10^{-6}$ (blue dashed), $A(T)=10^{-5}$ (blue dot-dashed), $A(T)=10^{-4}$ (blue dotted). These dephasings correspond to an underlying decoherence rate  of $\gamma_{\text{gdp}}= 10^{-4}g$ accumulated over each phasing gate of duration $T$. For a zero temperature Ohmic bath, the corresponding cuttoff frequencies are: $\omega_c/g=\{0.003,0.022,0.094\}$. Black line: performance without dynamical decoupling with $A_0(T)=0.0223$. e.g. if one were to switch the sign of the dispersive coupling during each GPG rather than flipping the spins. (b) Fidelity error for the same environments as above.}
\label{fig:relative_fidelity_supplementary}
\end{figure}

%\begin{figure}[tb]
%\begin{center}
%\includegraphics[width=8.6cm]{Figure6.pdf}
%\end{center}
%\caption{{\bf Suppression of dephasing via dynamical decoupling inherent in the sequence of GPGs used for each of the operators $U_s$ and $U_w$.} Solid curves are filter functions using the GPGs. Dashed curves are plots of Eq.~(\ref{filterapprox}), which is a good approximation for $\omega/g<1/\pi N$. Dot-dashed curves is the bare case without decoupling. Here (red, green, blue) curves correspond to $n=(10,100,1000)$ spins.}
%\label{fig:5}
%\end{figure}
In comparison, consider evolution where no spin flips are applied during the sequence, in which case the \emph{bare} functions are $C^{(0)}(t)=1\forall t\in[0,T)$, and $|f^{(0)}(\omega)|^2=4 \sin^2(T\omega/2)/\omega^2$. Results are plotted in Fig.~3 in the main paper and show there is substantial decoupling from the dephasing environment when the spectral density has dominant support in the range $\omega<g/2$. 
For $2\pi k \omega/g\ll1$, the summands in $f(\omega)$ can be expanded in a Taylor series in $\omega/g$ and to lowest order we find
\begin{equation}
g^2|f(\omega)|^2\approx \frac{(\omega/g)^2 \pi^4 N^2(N+2)^2}{9 (N+1)^2}.
\label{filterapprox}
\end{equation} 
This approximation is valid for $\omega/g <1/\pi N$, and, as shown in Fig. 3 in the main paper, for $1/\pi N<\omega/g< 1/2$ the function is essentially flat with an average value $g^2
 \overline{|f(\omega)|^2}\approx 3$ independent of $N$.  In the region $1/\pi N<\omega/g<1/2$ the bare filter function is oscillatory and has an average $g^2
 \overline{|f^{(0)}(\omega)|^2}\approx 13.63$, while for $\omega/g <1/\pi N$  it asymptotes to $ \frac{\pi^2 N^2(N+2)^2}{(N+1)^2}$. Thus, in the region $\omega/g<1/\pi N$ the ratio determining the reduction factor in the dephasing rate is $\frac{|f(\omega)|^2}{|f^{(0)}(\omega)|^2}=\pi^2\omega^2/g^2$, while for $\omega/g\in[1/\pi N,1/2]$, the reduction factor can be approximated by $\frac{\overline{|f(\omega)|^2}}{\overline{|f^{(0)}(\omega)|^2}}\approx 0.22$, provided the noise spectrum is sufficiently flat there. The effectiveness of this decoupling on precision and state precision fidelity is shown in Figure~\ref{fig:relative_fidelity_supplementary}.
 
Further, the aforementioned freedom to apply the GPGs in any order allows room for further improvement. For example, consider coupling to a zero temperature Ohmic bath with noise spectrum $S(\omega)=\alpha \omega e^{-\omega/\omega_c}$ and having cutoff frequency $\omega_c/g=0.1$. For $N=20$, the ratio of the effective decay rate for the linearly ordered sequence of GPGs above to that with no decoupling is $A(T)/A_0(T)=0.0085$. However, by sampling over permutations of the ordering of GPGs we find a sequence \cite{Note1}
%\footnote{The ordering $\{8,4,5,9,3,7,6,2,10,1\}$ achieves this. Note an exhaustive search over all $10!$ permutations was not done.}
achieving $A(T)/A_0(T)=0.0026$.  

\subsection{Preparation of error-tolerant states}

The state preparation method we have described so far has some inherent tolerance to decoherence. However, once the state is prepared, further errors could accumulate such as qubit loss or dephasing, while waiting for the accumulation of the measurement signal. Some strategies to address this were recently proposed in Ref. \cite{ouyang2019} where they suggest using superpositions of Dicke states as probe states. 
The class of states considered there are
\[
\ket{\varphi_u}=\frac{1}{\sqrt{2^n}}\sum_{j=0}^n\sqrt{\binom{n}{j}}\ket{J=\frac{knu}{2},M=kj-J}.
\]
Here the number of spins $N=k\times n\times u$, and the parameters $u$ and $n$ determine the robustness of the states to some number of loss and dephasing errors respectively, while $k$ is a parameter to scale the number of qubits in the superposition (larger $k$ means better performance). The case $u=1$ tolerates erasure errors; specifically, the state $\ket{\varphi_1}$ has a large quantum Fisher information obeying Heisenberg scaling when the number of erasure errors is less than $n$. 
We will consider the state performing well in the presence of one erasure error:  $u=1, n=2, k=N/2$, which can be written
\begin{equation}
\ket{\varphi_1}=\frac{1}{2}(\ket{J,-J}+\sqrt{2}\ket{J,0}+\ket{J,J}).
\end{equation}
The case $u=2$ tolerates a constant number of dephasing errors. We will focus on the state with $u=2,n=1,k=N/2$ which tolerates one dephasing error and can be written
\begin{equation}
\ket{\varphi_2}=\frac{1}{\sqrt{2}}(\ket{J,-J}+\ket{J,0}).
\end{equation}

We now describe how to make these states using our protocol.
% The state $\ket{J,0}$ can be prepared starting in the spin coherent state $\ket{s}=e^{i\frac{\pi}{2}J^y}\ket{J,-J}$, by applying a sequence of $\# G=O(N^{1/4})$ Grover steps where the
% Grover unitary is the composition
% $U_G=U_s U_w$. Here $U_w=W(N/2)$ and $U_s=e^{iJ^y \pi/2}W(0)e^{-iJ^y\pi/2}$ where the operator $W(\ell)=e^{i\pi \ket{J,\ell-N/2}\bra{J,\ell-N/2}}$, applies a phase to a single Dicke state. Both $U_s$ and $U_w$ are generated by $N/2$ applications of GPGs with varying parameters. The total number of GPGs needed for each Grover step is $N$.
A key ingredient to prepare a superposition of Dicke states is to perform a controlled state preparation. If we introduce an ancilla spin $A$ which can be allowed to couple to the mode when the other spins do not (e.g. by detuning the other spins far away from the dispersive coupling regime), then a controlled displacements of the mode can be done:
\begin{equation}
\begin{array}{lll}
\Lambda(\beta)&=&\ket{0}_A\bra{0}\otimes {\bf 1}+\ket{1}_A\bra{1}\otimes D(\beta)\\
&=&D(\beta/2)R(\pi \ket{1}_A\bra{1}) D(-\beta/2)R(-\pi\ket{1}_A\bra{1}).
\end{array}
\label{contrdis}
\end{equation}
Here $R(\pi \ket{1}_A\bra{1})=e^{i \pi a^{\dagger}a \ket{1}_A\bra{1}}$, meaning only the ancilla state $\ket{1}_A$ couples to the mode.
Now replacing the displacements $D(\beta)$ and $D(-\beta)$ with the controlled displacements $\Lambda(\beta)$ and $\Lambda(-\beta)$, the effect is a controlled GPG (see also Ref \cite{brennen2016}):
\begin{equation}
\Lambda(U_{\rm GPG})=\ket{0}_A\bra{0}\otimes {\bf 1}+\ket{1}_A\bra{1}\otimes U_{\rm GPG}.
\end{equation}
Thus by simply replacing all instances of GPGs with controlled GPS we can achieve a controlled Grover step unitary $G$. Note the unitary $e^{iJ^y \pi/2}$ conjugating $W(0)$ in $U_s$ does not need to be controlled, meaning the entire unitary $U_G^{\#G}$ can be made into a controlled unitary
\begin{equation}
\Lambda(U_G^{\# G})=(\ket{0}_A\bra{0}\otimes {\bf 1}+\ket{1}_A\bra{1}\otimes U_G)^{\#G}.
\end{equation}
This is not quite enough. The state preparation of a Dicke states described above applies $U_G^{\# G}$ to a particular initial state, namely the spin coherent state $\ket{s}$. We will also require a way to perform a controlled rotation on all the spins of the form
\begin{equation}
\Lambda( e^{i J^y\pi/2})=\ket{0}_A\bra{0}\otimes {\bf 1}+\ket{1}_A\bra{1}\otimes e^{i J^y\pi/2}.
\end{equation}
Without having direction interactions between the ancilla and the other spins it is not obvious how to do this. However, it is possible to mediate the interaction with the mode by choosing $\phi=0$ and $\theta\ll 1$ in one instance of a controlled GPG. This will give $\Lambda(U_{\rm GPG}(\theta,0,\pi/4\theta)\approx \ket{0}_A\bra{0}\otimes {\bf 1}+\ket{1}_A\bra{1}\otimes e^{-i J^z\pi/2}$ where we have approximated $\sin(\theta J^z)\approx \theta J^z$. Note, in order for this to be valid we require $\theta \ll 1/N$ and consequently $\chi=|\alpha|^2\gg N$, i.e. the area of the GPG in phase space needs to grow with $N$, or the gate could be composed into $N$ GPGs each of area of $O(1)$. This will consequently incur a loss of fidelity due to mode decay
\begin{equation}
F_{\mathrm{pro}}(\mathcal{E},U_{\rm GPG})>e^{-6\pi|\alpha|^2 \kappa/g}\cos(|\alpha|^24\pi \kappa/g).
\end{equation}
but no worse than the performance for state preparation without ancilla. 

The controlled operation is then
\[
\Lambda( e^{i J^y\pi/2})=e^{-i J^x\pi/2 } \Lambda(U_{\rm GPG}(\theta,0,\pi/4\theta) e^{i  J^x \pi/2}.
\]

We can now write the process to prepare the state $\ket{\varphi_2}$:
\begin{enumerate}
\item
Prepare the product state $\frac{1}{\sqrt{2}}(\ket{0}+\ket{1})_A\otimes \ket{J,-J}$.
\item
Apply $e^{-i J^x \pi/2 } \Lambda(U_{\rm GPG}(\theta,0,\pi/4\theta) e^{i  J^x \pi/2}$.
\item
Apply $\Lambda(U_G^{\# G})$. This involves $N\times \#G$ instances of $\Lambda(U_{\rm GPG})$ for varying parameters.
\item
Measure the ancilla in the $\ket{\pm_x}_A$ basis. The outcomes $r=\pm 1$ each occur with probability $1/2$. 
The conditional system state is
\[
\frac{1}{\sqrt{2}}(\ket{J,-J}\pm \ket{J,0})
\]
\item
Apply the classically controlled product unitary $Z(r)={Z^{(1-r)/2}}^{\otimes N}$. If we assume $N/2$ is odd then
\[
Z(r)\frac{1}{\sqrt{2}}(\ket{J,-J}\pm \ket{J,0})=\ket{\varphi_2}.
\]
\end{enumerate}

To prepare the state $\ket{\varphi_1}$ a similar process can be used. However, rather than the $\ket{0}_A$ state being correlated with the product state $\ket{J,-J}$ we want it correlated with the GHZ state $\frac{1}{\sqrt{2}}(\ket{J,-J}+\ket{J,J})$. Such a state can be prepared using one additional controlled GPG gate. This follows from the observation that $e^{i\frac{J^y\pi}{2}} U_{\rm GPG}(\pi,\pi/2,\frac{\pi}{8})e^{-i\frac{J^y\pi}{2}}\ket{J,-J}=\frac{1}{\sqrt{2}}(\ket{J,-J}+\ket{J,J})$.
These two processes are summarized by the following circuits:
\begin{widetext}

\[
\Qcircuit @C=.5em @R=.2em @!R 
{
\lstick{\ket{0}_A} & \gate{H} & \ctrl{1} & \qw &  \ctrl{1} & \gate{H} &\meter \cwx[1] \\
\lstick{\ket{0}^{\otimes N}} & \gate{e^{i\frac{J^x\pi}{2}}} & \gate{U_{\rm GPG}(\theta,0,\frac{\pi}{2\theta})} & \gate{e^{-i\frac{J^x\pi}{2}}} &  \gate{U_G^{\# G}} & \qw & \gate{Z^{\otimes N}}
 & \rstick{\ket{\varphi_2}} \qw 
}
\]

\[
\Qcircuit @C=.5em @R=0em @!R  
{
\lstick{\ket{0}_A} & \gate{HX} & \qw & \ctrl{1} & \gate{X} & \ctrl{1} & \qw & \ctrl{1} & \gate{H} &\meter  \cwx[1] \\
\lstick{\ket{0}^{\otimes N}} & \gate{e^{i\frac{J^y\pi}{2}}} & \qw & \gate{U_{\rm GPG}(\pi,\pi/2,\frac{\pi}{8})} & \gate{e^{-i\frac{J^y\pi}{2}}e^{-i \frac{J^x\pi}{2}}} & \gate{U_{\rm GPG}(\theta,0,\frac{\pi}{4\theta})} & \gate{e^{i\frac{J^x\pi}{2}}} &  \gate{U_G^{\# G}} & \qw & \gate{Z^{\otimes N}} & \rstick{\ket{\varphi_1}} \qw 
}
\]
\end{widetext}

\end{document}